\documentclass[fleqn,usenatbib]{mnras}

\usepackage[dvipsnames]{xcolor}
\usepackage{tikz,hyperref}
\usepackage[normalem]{ulem}

\definecolor{lime}{HTML}{A6CE39}
\DeclareRobustCommand{\orcidicon}{
	\begin{tikzpicture}
	\draw[lime, fill=lime] (0,0) 
	circle [radius=0.13] 
	node[white] {{\fontfamily{qag}\selectfont \tiny ID}};
	\draw[white, fill=white] (-0.0625,0.095) 
	circle [radius=0.007];
	\end{tikzpicture}
	\hspace{-2mm}
}

\foreach \x in {A, ..., Z}{\expandafter\xdef\csname orcid\x\endcsname{\noexpand\href{https://orcid.org/\csname orcidauthor\x\endcsname}
			{\noexpand\orcidicon}}
}

\usepackage{newtxtext,newtxmath}

\usepackage[T1]{fontenc}

\DeclareRobustCommand{\VAN}[3]{#2}
\let\VANthebibliography\thebibliography
\def\thebibliography{\DeclareRobustCommand{\VAN}[3]{##3}\VANthebibliography}

\usepackage{graphicx}	
\usepackage{amsmath}	

\title[A morpho-kinematic view of the Necklace nebula]{A morpho-kinematic view of the Necklace nebula}

\author[E.\ Santamar\'\i a et al.]{E.~Santamar\'{i}a$^{1,2}\orcidA$, 
J.~A.~Toal\'{a}\thanks{E-mail:\,j.toala@irya.unam.mx}$^{1}\orcidB,$
L.~Sabin$^{3}\orcidF,$
M.~A.~Guerrero$^{4}\orcidC$ ~and
G.~Ramos-Larios$^{2}\orcidD$
\\
$^{1}$Instituto de Radioastronomía y Astrofísica, Universidad Nacional Autónoma de México, Morelia 58089, Michoacán, Mexico\\
$^{2}$Instituto de Astronom\'{i}a y Meteorolog\'{i}a, Dpto. de Física, CUCEI, Universidad de Guadalajara, Guadalajara, Jalisco, Mexico\\
$^{3}$Instituto de Astronomía, Universidad Nacional Autónoma de México, AP 106,  Ensenada 22800, BC, Mexico\\
$^{4}$Instituto de Astrof\'{i}sica de Andalucía, CSIC, Glorieta de la Astronomía S/N, Granada 18008, Spain
}

\date{\today}

\pubyear{\the\year{}}

\begin{document}
\label{firstpage}
\pagerange{\pageref{firstpage}--\pageref{lastpage}}
\maketitle

\begin{abstract}
We present a morpho-kinematic study of the Necklace Nebula based on spatially resolved spectroscopic observations and 3D modeling performed with the {\sc shape} software. 
The nebula is characterized by an equatorial disrupted ring accompanied by bipolar outflows. 
The geometry and velocity field of these different components are derived from the comparison between observed and synthetic images and position-velocity diagrams. 
The best model successfully reproduces the main morphological and kinematic features of the nebula, allowing us to constrain its 3D structure. We confirm that the disrupted ring is tilted about 60$^\circ$ from the plane of the sky, but the bipolar outflows are inclined by $46^{\circ}$ with respect to the line of sight. 
This situation might indicate that the bipolar structures are not simple extensions of the equatorial ring, but instead correspond to a distinct mass-ejection event. 
The slight misalignment between these two components suggests either precession of the ejection axis, a temporal change in the dominant mass-loss geometry, or a misalignment in the magnetic field. Given that the Necklace Nebula hosts a close binary central star and is a well-established post-common envelope planetary nebula, our results support a scenario in which binary interaction plays a central role in shaping its morphology, even before the common envelope phase. 
\end{abstract}



\section{Introduction}

The origin of collimated outflows in planetary nebulae (PNe) remains an open problem closely linked to binary interaction and mass accretion processes \citep{Morris,Soker}. In post-common-envelope (CE) systems, the lack of observational evidence for ongoing accretion discs complicates the identification of the jet-launching mechanism. Nevertheless, kinematic studies have shown that in several post-CE PNe the jets are older than the main nebular shell, implying that they were launched prior to the CE phase \citep{Mitchell,Corradi11,Miszalski11,2012Boffin,Tocknell2014}. A decisive observational breakthrough was the detection of a carbon-enriched main-sequence companion ($C/O > 1$) in the central star of the Necklace Nebula \citep[a.k.a. PN G~054.2$-$03.4 and IPHASX J194359.5+170901;][]{2008PhS,Viironen,Corradi11} by \citet{Miszalski13}. This finding provides the first direct proof of efficient accretion before the CE phase and strongly supports a scenario in which jets are launched from transient accretion discs formed through wind accretion during the early AGB phase. However, the apparent lack of carbon enrichment in the inner nebula remains puzzling, as it contrasts with the carbon-rich nature of the companion and challenges a straightforward evolutionary interpretation \citep{Jones2026}.

\begin{figure*}
    \centering
	\includegraphics[width=1.0\linewidth]{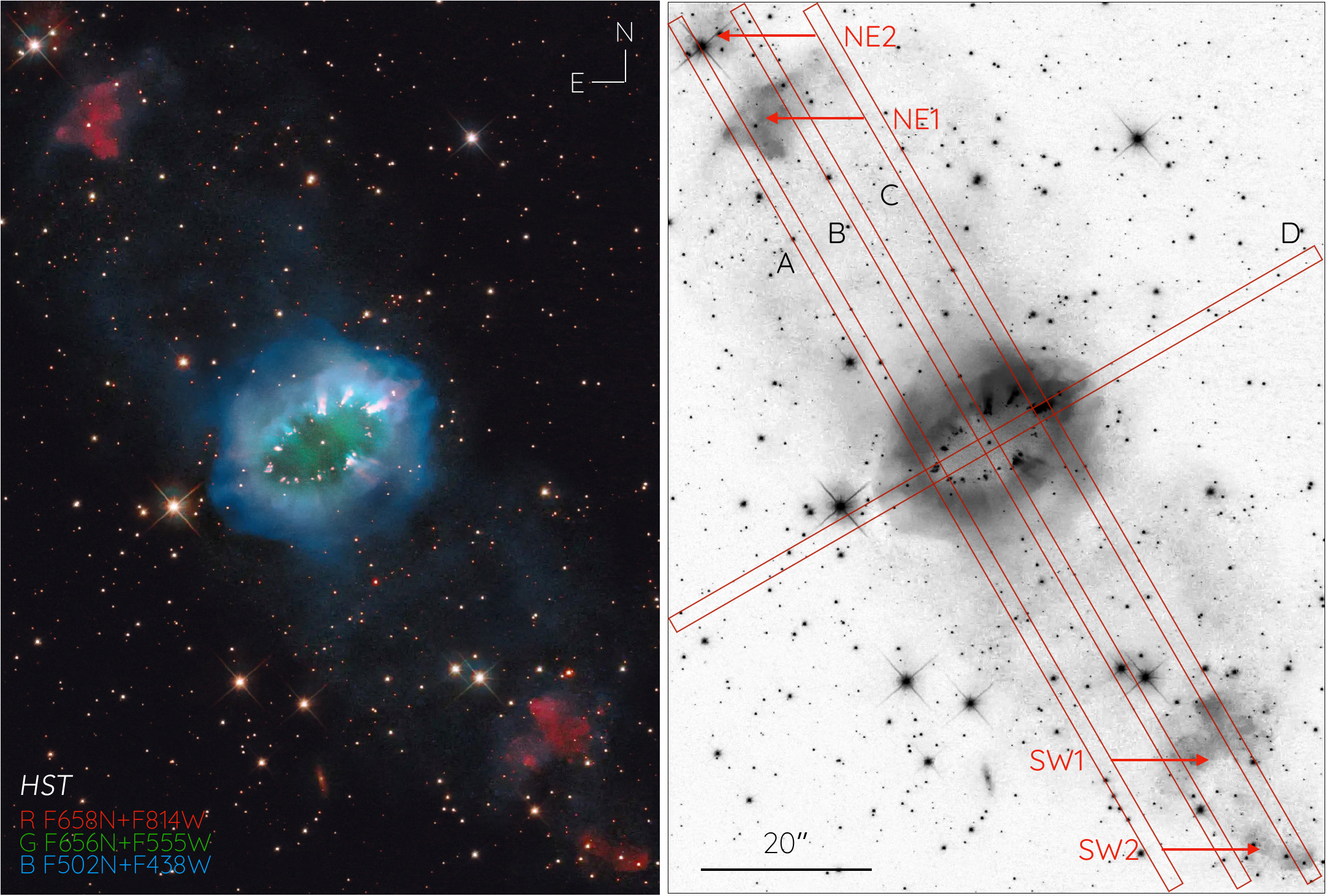}
    \caption{Optical images of the Necklace Nebula obtained with {\it HST} data. (Left panel:) Color-composite image where blue is F658N+F814W filters, green corresponds to F656N+F555W, and blue is F502N+F438W. (Right panel:) Same as left panel but in gray-scale. The slits positions corresponding to the SPM MES observations are overplotted in red color. The knots denoted as NE 1-2 and SW 1-2 are marked with red arrows.}
    \label{fig:slits}
\end{figure*}

The Necklace Nebula was discovered as part of the Isaac Newton Telescope Photometric H$\alpha$ Survey of the Northern
Galactic Plane \citep{Drew2005}. The IPHAS survey allowed the discovery of very compact as well as extended (evolved) PNe at the faint end of the PNe luminosity function \citep[e.g.,][]{Viironen2009,Sabin2014}. Within IPHAS, the Necklace Nebula stood out as one of the brightest PNe. Narrow-band imaging presented by \citet{Corradi11} showed a central ring-like structure with a collection of low-ionization clumps with outward faint tails. It also exhibits a bipolar morphology unveiled by the presence of low-ionization structures -- apparently aligned perpendicularly to the ring-like structure -- and a pair of cylindrical cavities. The general orientation in the plane of the sky of the Necklace Nebula is in the NE-SW direction with an approximate position angle (PA) of 34$^\circ$ (see Fig.~\ref{fig:slits}).

\citet{Corradi11} used medium-resolution Nordic Optical Telescope spectra 
to assess the Doppler shifted [N~{\sc ii}] emission of the clumps in the ring-like structure to peer into the kinematics of the Necklace Nebula. They estimated a radius of $r=6.5\pm0.5$~arcsec, inclination of $59\pm3^\circ$ (where 0$^\circ$ corresponds to a ring in the plane of the sky, while 90$^\circ$ would be an edge-on ring), and an expansion velocity of $28\pm3$~km~s$^{-1}$. At a distance of 4.6 kpc\footnote{See \cite{Jones2026} for an extensive and updated discussion on the distance towards the Necklace Nebula. Note that our adopted distance of 4.6~kpc \citep{Corradi11} is in good agreement, within the uncertainties, with the {\it Gaia} parallax-based distance favored by \cite{Jones2026}.} the estimated kinematical age of this structure would be $\tau_\mathrm{k} \simeq 5060\pm500$~yr. Under the assumption that the low-ionization structures corresponding to the jets in the NE-SW extension of the Necklace nebula are aligned perpendicularly to the ring, \citet{Corradi11} estimate their age to be twice as that of the central ring structure.

The growing body of hydrodynamical simulations suggest that the Necklace Nebula can naturally be explained as the result from the post–CE evolution of a binary system \citep[see][and references therein]{GarciaSegura2018,GarciaSegura2020,Ondratschek2022,Vetter2025}. A toroidal dense equatorial structure is formed around the binary while jets shape the elongated structures perpendicularly towards the poles. The disruption of the toroidal structure might be related to the subsequent action of photoionization from the post-CE WD and the very likely developing of a fast wind. The combination of effects disrupted the dense material of the toroidal structure forming the fragmented ring-like structure as illustrated by the simulations presented in, e.g., \citet{GarciaSegura2022}.

In this work, we present high-dispersion spectroscopic observations in combination with high-quality narrow-band images to peer into the details of the morpho-kinematics of the Necklace Nebula. The paper is organized as follows. In Section~\ref{sec:obs} we describe the observations and their analysis. In Section~\ref{sec:shape} we propose a morpho-kinematic model of the different structures detected in the Necklace Nebula and estimate their dynamical ages. The discussion and conclusions are presented in Section~\ref{sec:summary}.

\section{Observations and data preparation}
\label{sec:obs}

\subsection{{\it Hubble Space Telescope} images}

The Necklace Nebula was observed by the {\it Hubble Space Telescope (HST)} on 2011 July 2 as part of proposal 12675 (PI: Keith Noll). The observations were taken with the Wide Field Camera 3 (WFC3) in the UV/visible (UVIS) channel, through the F502N, F656N, and F658N narrow-band filters that include the emission from [O~{\sc iii}]~5007~\AA, H$\alpha$, and [N~{\sc ii}]~6584~\AA, respectively. In addition, images with the F438W, F555W, and F814W broad-band filters were also obtained. The processed images were retrieved from the Hubble Legacy Archive\footnote{\url{https://hla.stsci.edu/}}. 

We used the {\it HST} data to produce the color-composite image of Fig.~\ref{fig:slits}-left\footnote{A similar image has been presented by the Hubble Mission Team in \url{https://science.nasa.gov/centers-and-facilities/goddard/hubble-views-a-dazzling-cosmic-necklace/}}. Each channel emission was produced by combining a narrow-band filter with one of the broad-band images. That is, the red channel is the combination of data from the F658N and F814W filters. The green channel is the result of combining the filters F656N and F555W. Finally, the blue channel in the figure corresponds to F502N and F438N. For comparison and further discussion, the right panel of Fig.~\ref{fig:slits} presents the same figure as that from the left panel but in gray-scale. 

The RGB image in Fig.~\ref{fig:slits}-left reveals the intricate complexity of the Necklace Nebula. Notably, it highlights the double-clump morphology of the low-ionization structures located at the extremities of the NE-SW axis. The SW clumps consist of at least two distinct structures, labeled here as SW1 and SW2 (see Fig.~\ref{fig:slits}, right). The low-ionization clump NE1 appears to be the morphological counterpart of SW1. Furthermore, another clump, NE2 -- projected in close proximity to a bright star -- is clearly unveiled in the gray-scale image.

\subsection{High-dispersion spectroscopy}

Long-slit, high resolution spectroscopic observations of the Necklace Nebula were obtained using the Manchester Echelle Spectrometer \citep[MES;][]{Meaburn} mounted on the 2.1 m telescope at the Observatorio Astron\'{o}mico Nacional in San Pedro M\'{a}rtir (SPM, Ensenada, Mexico)\footnote{The Observatorio Astronómico Nacional at the Sierra de San Pedro Mártir (OAN-SPM) is operated by the Instituto de Astronomía of the Universidad Nacional Autónoma de México.}. The SPM MES observations of the Necklace nebula were obtained on 2025 June 3. A summary of the observations is presented in Table~\ref{tab:obs} and the slit positions are shown in Fig.~\ref{fig:slits}-right. The slit positions are labeled as A, B, C, and D.

The E2V 42-40 CCD, with a pixel size of 13.5 $\mu$m was used with 4$\times$4 on-chip binning, which resulted in a plate scale of 0.702 arcsec pix$^{-1}$. Observations were made in the spectral range of H$\alpha$ and [N~{\sc ii}] emission lines using a filter ($\Delta \lambda=90$ \AA) to isolate the 87th echelle order (0.10 \AA~pix$^{-1}$ spectral scale). We also obtained spectra with the interference filter $\lambda_{\rm c}=5020$ \AA  ~($\Delta \lambda=70$ \AA) that includes the emission line of [O\,{\sc iii}] $\lambda5007$ \AA ~to isolate the 114th echelle order (0.086 \AA~pix$^{-1}$ spectral scale). 
All observations adopted a slit width of 150 $\mu$m  (1.9 arcsec), resulting in a spectral resolution of $12\pm1$ km s$^{-1}$.

The spectra were processed using standard calibration routines in {\sc iraf} \citep{Tody}\footnote{{\sc iraf} is distributed by the National Optical Astronomy Observatory, which is operated by the Association of Universities for Research in Astronomy (AURA) under a cooperative agreement with the National Science Foundation.}, including bias subtraction and wavelength calibration with ThAr arc lamps obtained immediately before and after the science observations. The wavelength accuracy is estimated to be $\pm$1 km s$^{-1}$. As determined from the FWHM of the continuum emission of stars registered along the slit, the spatial resolution (seeing) is estimated to be $\simeq1.7$~arcsec

\begin{figure*}
    \centering
	\includegraphics[width=0.8\linewidth]{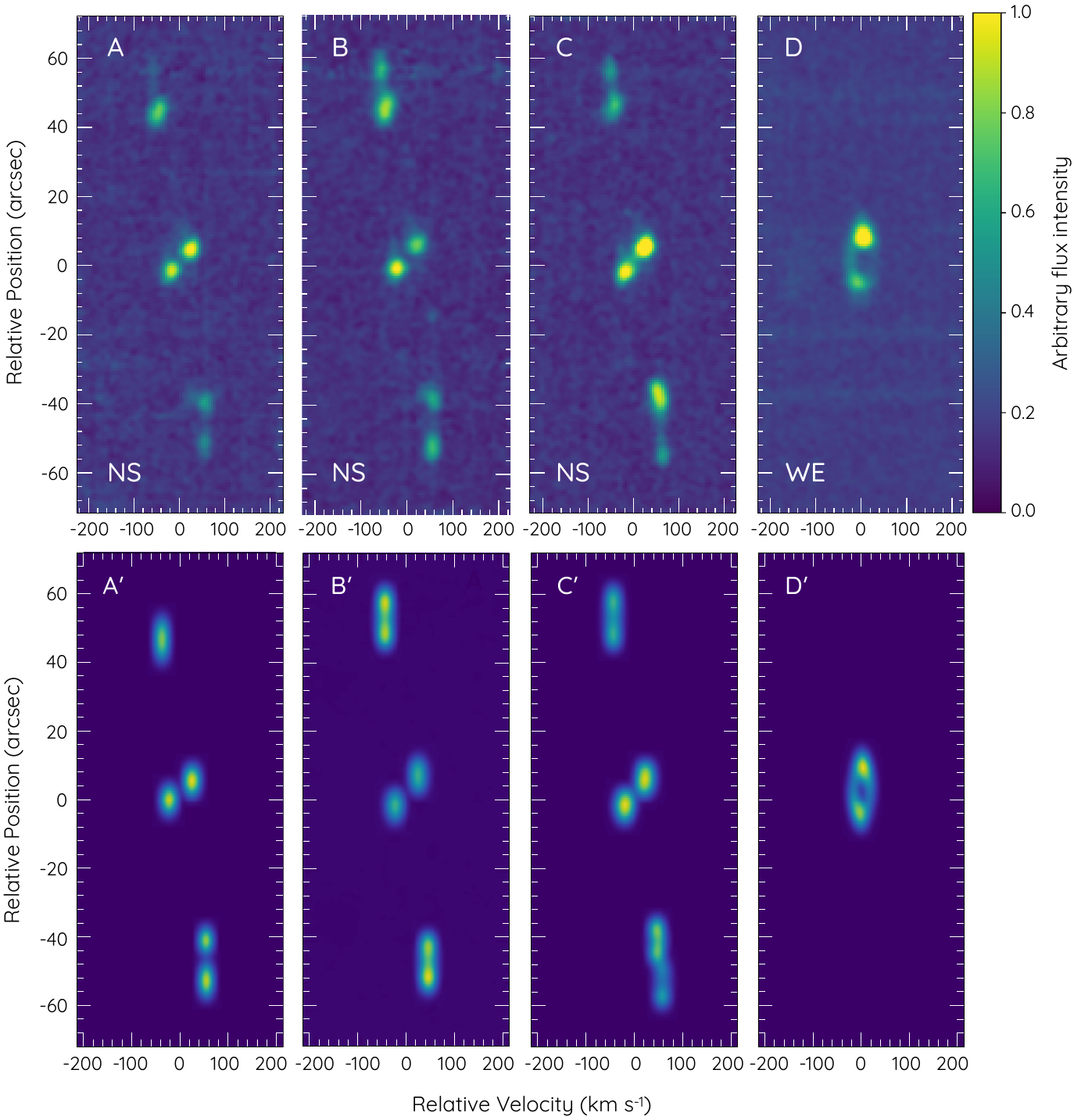}
    \caption{Top: PV diagrams of slits A, B, C, and D extracted from the [N~{\sc ii}] $\lambda6584$~\AA~ emission line detected from the Necklace Nebula. Bottom: Synthetic PV diagrams obtained from our best {\sc shape} morpho-kinematic model.}
    \label{fig:N2}
\end{figure*}

\begin{figure*}
    \centering
	\includegraphics[width=0.8\linewidth]{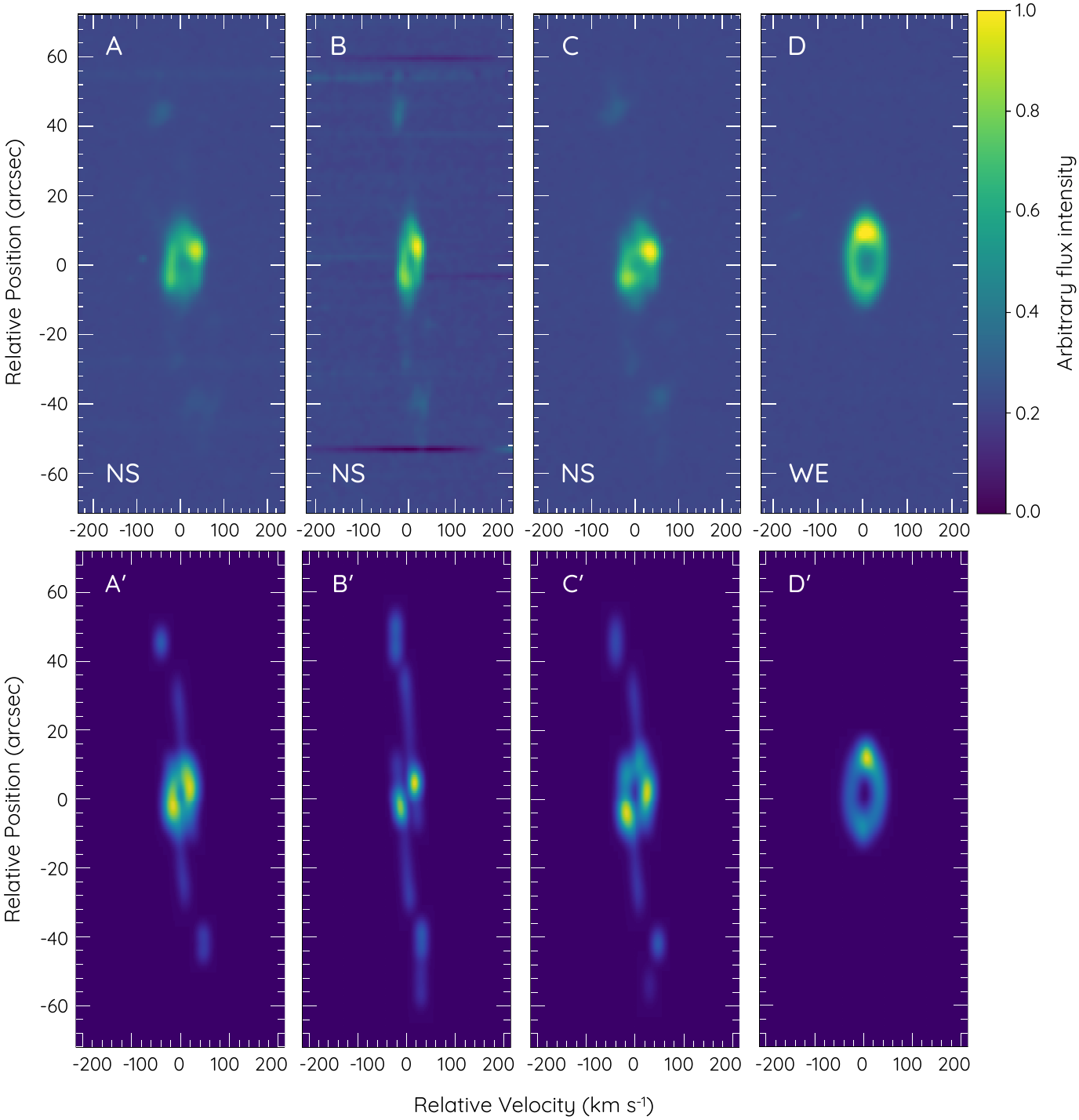}
    \caption{Top: PV diagrams of slits A, B, C, and D extracted from the [O~{\sc iii}] $\lambda5007$~\AA~ emission line detected from the Necklace Nebula. Slits A, B, C, and D are illustrated in the middle panel of Fig.~\ref{fig:slits}. Bottom: Synthetic PV diagrams obtained from our best {\sc shape} morpho-kinematic model.}
    \label{fig:O}
\end{figure*}

The position velocity (PV) diagrams extracted from slits A, B, C, and D are displayed in the top rows of Figs.~\ref{fig:N2} and \ref{fig:O}, corresponding to the [N\,{\sc ii}] $\lambda6584$ \AA ~and [O\,{\sc iii}] $\lambda5007$ \AA ~emission lines, respectively. We note that the PV structure of H$\alpha$ closely resembles that of [N\,{\sc ii}] $\lambda6584$ \AA, although the former is broadened thermally. Complementary PV diagrams of the H$\alpha$ and [N~{\sc ii}]~$\lambda6548$~\AA~ are provided in Appendix~\ref{app:PV} (see Fig.~\ref{fig:N1}).
All radial velocities were corrected to the Local Standard of Rest. From these corrected data, we derive a systemic velocity of $V_{\rm{sys}} = 43\pm3~\rm{km}~{s}^{-1}$. This measurement is in excellent agreement with the value of $45\pm2~\rm{km}~{s}^{-1}$ previously reported by \cite{Corradi11}.

\begin{table}
\caption{Details of the SPM-MES observations of the Necklace nebula. The offsets are denoted with respect to the position of the central star.}
\label{slits}
\centering
\setlength{\tabcolsep}{\tabcolsep}   
\label{tab:obs}
\begin{tabular}{cccrcc} 
\hline
Slits &   Offset  &  Filter   & PA    & Binning            & Exp. Time \\
      &   (arcsec)  &           &($^{\circ}$) &             & (s)     \\
\hline
A & $-$7.0 & H$\alpha+$[N\,{\sc ii}] & $+$32 & 4$\times$4  & 1200  \\
A & $-$7.0 & [O\,{\sc iii}] & $+$32 & 4$\times$4  & 1200  \\
B & 0.0 & H$\alpha+$[N\,{\sc ii}] & $+$32 & 4$\times$4  & 1200  \\
B & 0.0 & [O\,{\sc iii}] & $+$32 & 4$\times$4  & 1200  \\
C & $+$7.0 & H$\alpha+$[N\,{\sc ii}] & $+$32 & 4$\times$4  & 1200  \\
C & $+$7.0 & [O\,{\sc iii}] & $+$32 & 4$\times$4  & 1200  \\
D & 0.0 & H$\alpha+$[N\,{\sc ii}] & $-$58 & 4$\times$4  & 1200  \\
D & 0.0 & [O\,{\sc iii}] & $-$58 & 4$\times$4  & 1200  \\
\hline
\end{tabular}
\end{table}

\begin{table}
\caption{
Kinematical properties of the Necklace nebula. 
All expansion velocities ($V_{\rm exp}$) and angular distance to the CSPN ($r$) are de-projected values derived from the morpho-kinematic model. 
These values are used to compute the kinematical age ($\tau_\mathrm{k}$), where a distance of $d=4.6$ kpc has been adopted. For the extended bipolar structure, single values for $V_{\rm exp}$ and $r$ are omitted; its kinematics are instead governed by the homologous expansion law described in Eq.~(\ref{eq:bipolar}).
For the outer knots, two values of the inclination are assumed, that of the ring (60$^{\circ}$) and that of the bipolar structure ($46^{\circ}$).
}
\label{tab:kin}
\centering
\setlength{\tabcolsep}{\tabcolsep}   
\label{tab:kin}
\begin{tabular}{lcccc} 
\hline
Structure &  $V_{\rm exp}$  & $r$ & $i$ & $\tau_{\rm k}$  \\
 &   (km s$^{-1}$)  & (arcsec)   & ($^{\circ}$) & (yr)  \\
\hline
Ring & $30\pm5$ & $7\pm1$ & $60^{\circ}$ & $5,090\pm1,900$ \\
Bipolar & -- & -- & $46^{\circ}$ & $18,200\pm2,200$ \\
\hline
NE1 & $83\pm7$ & $54\pm2$ & $60^{\circ}$ & $14000\pm1800$ \\
NE2 & $85\pm7$ & $61\pm2$ & $60^{\circ}$ & $15300\pm1800$ \\
SW1 & $96\pm7$ & $54\pm2$ & $60^{\circ}$ & $11900\pm1300$ \\
SW2 & $99\pm7$ & $62\pm2$ & $60^{\circ}$ & $13500\pm1400$ \\
\hline
NE1 & $58\pm5$ & $67\pm2$ & $46^{\circ}$ & $25,200\pm3,200$ \\
NE2 & $60\pm5$ & $76\pm2$ & $46^{\circ}$ & $27,600\pm3,300$ \\
SW1 & $68\pm5$ & $67\pm2$ & $46^{\circ}$ & $21,500\pm2,400$ \\
SW2 & $70\pm5$ & $78\pm2$ & $46^{\circ}$ & $24,300\pm2,500$ \\
\hline
\end{tabular}
\end{table}

\section{Unveiling the 3D structure of the Necklace}
\label{sec:shape}

To investigate the 3D morpho-kinematic structure of the Necklace Nebula, we employ the software {\sc shape} \citep{Steffen2011}. 
It enables the construction of predefined geometrical components with prescribed velocity fields that are used to generate synthetic images and spectra. 
These can be directly compared with the observations and the geometrical components interactively modified (for example, by rotating components, adjusting their spatial extents and expansion velocities) to reduce the differences between the synthetic and  observed images and spectra. 
This iterative approach allows converging towards a model that simultaneously reproduces multiple morphological and kinematic observables, providing a robust representation of the nebula’s 3D structure and expansion patterns. 
{\sc shape} has been used to address the morpho-kinematic structures around other evolved stars  \citep[see, e.g.,][and references therein]{GomezMunoz2023,RodriguezGonzalez2022,Santamaria2024,Zavala2022}.

It is important to highlight that {\sc shape} does not model the physical gas density. Instead, the point density parameters are used to regulate and enhance the emission of the user-defined structures. Accordingly, reproducing the observed brightness distribution requires the introduction of structure-dependent density variations within the model. As a result, our {\sc shape} model is intended solely to assess the kinematic properties of the structures and should not be interpreted in terms of physical density.

A single global systemic velocity of $V_{\rm sys} = +43 \pm 3 \text{ km s}^{-1}$ (LSR) was adopted for the entire morpho-kinematic model. 
This value was strictly anchored by fitting the kinematic center of the main equatorial ring, which is the brightest and most structurally coherent component of the nebula. Altering $V_{\rm sys}$ to minimize the velocity differences between opposing polar knots critically degrades the highly symmetric fit of the equatorial ring. 

\begin{figure*}
    \centering
	\includegraphics[width=1.0\linewidth]{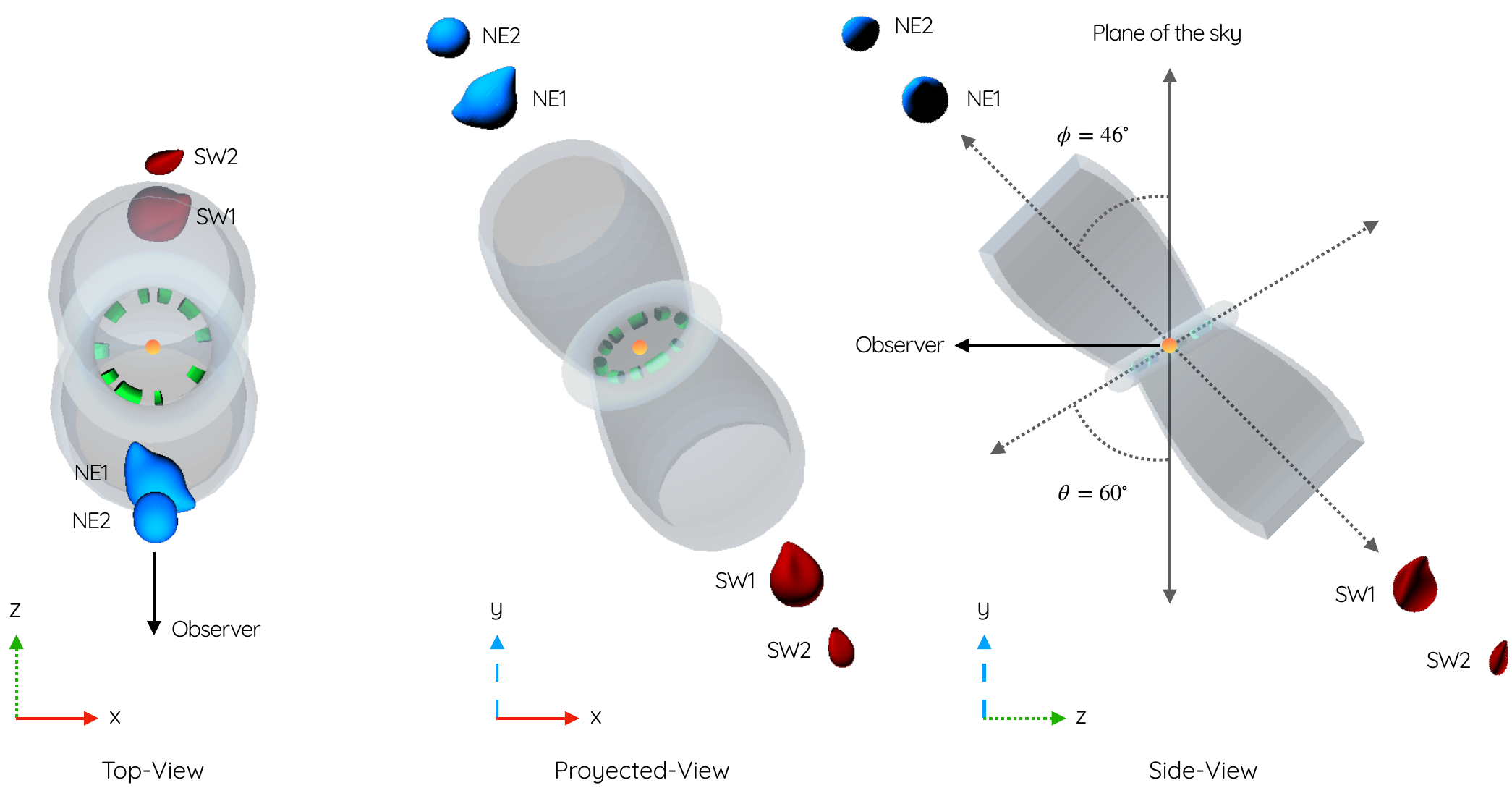}
    \caption{Morphological components used in our best {\sc shape} model of the Necklace Nebula. This model was used to extract synthetic PV diagrams shown in the bottom rows of Fig.~\ref{fig:N2} and \ref{fig:O}. The top-view (left) represents the nebula as seen through the fragmented ring projected in the $z-x$ plane. The central panel shows the structures in the plane of the sky of the observation (the $y-x$ direction). The side-view (right) corresponds to the $y-z$ direction, which is not aligned with the plane of the sky, but allows the inclination of the structures with respect to the plane of the sky to be clearly assessed. In the {\sc shape} model, the gray shaded torus represents the continuous geometric and kinematic boundary of the equatorial outflow seen in the [O~{\sc iii}] emission line. Whereas the green broken ring and blue and red knots correspond to the localized, high-density clumps which provide the dominant emission contribution in the [N~{\sc ii}] emission line.}
    \label{fig:MO}
\end{figure*}

We started the {\sc shape} model assessing the properties of the brightest region in the Necklace Nebula, its central disrupted ring. 
As a first approach we used a ring structure to estimate the best inclination angle\footnote{
We adopt the notation of \citet{Corradi11}, where an inclination of 0$^\circ$ corresponds to a ring on the plane of the sky, and an inclination of 90$^\circ$ corresponds to an edge-on ring.
}. 
To refine our model, the ring structure has been segmented into ten sections to best fit the {\it HST} images (see features in green in Fig.~\ref{fig:MO}). 
We confirm \citet{Corradi11}'s finding that the ring has an inclination with respect to the plane of the sky of $60^\circ\pm2^\circ$. Its projected major axis has a PA of 122$^\circ$ with an angular radius of 7$\pm$1 arcsec, which at the distance of 4.6 kpc implies a physical radius of 0.16$\pm$0.07~pc. 
The true expansion velocity of the ring resulted by the model is 30$\pm$5~km~s$^{-1}$, thus the clumps of the necklace-like structure have a kinematic age  $\tau_\mathrm{k,ring}=5090\pm1900$~yr. Details of the properties of the ring-like structure are listed in Table~\ref{tab:kin}.

As for the bipolar structure, it was initially assumed to consists of cylindrical components emerging orthogonally from the ring (i.e.\ to the plane containing the clumps) following the suggestion of \citet{Corradi11}. 
However, this configuration produced synthetic PV diagrams that failed to reproduce the observed ones. 
A significantly better fit was obtained by 
adopting an inclination angle of the major axis of the cylinder with respect to the line of sight of $46^\circ\pm2^\circ$ (see Fig.~\ref{fig:MO}, side-view). 
The major axis of this bipolar structure is tilted by 14$^\circ$ with respect to an axis orthogonal to the ring of low-ionization knots. 
Moreover it is found that this structure is not strictly a cylinder, but the best model requires it to have an hourglass-like structure as illustrated in Fig.~\ref{fig:MO}. 

The extended bipolar structure was then modeled assuming a homologous expansion law, where the deprojected velocity is proportional to the distance from the central star as 
\begin{equation}
    V = (8.4\pm0.9) \left( \frac{r}{7~\mathrm{arcsec}}\right)~\mathrm{km}~\mathrm{s}^{-1}.
\label{eq:bipolar}
\end{equation}
\noindent 
The velocity gradient is thus $1.20 \pm 0.12 \text{ km s}^{-1} \text{ arcsec}^{-1}$. 
At its equator, at the shared plane of the ring, the bipolar structure has a radius of 7~arcsec and therefore an expansion velocity of 8.4 km~s$^{-1}$. 
At its outermost point along the symmetry axis, which lies at $46\pm2$~arcsec from the CSPN, the expansion velocity is $55\pm5$ km~s$^{-1}$. 
Using this gradient and its associated uncertainties, the kinematic age for the bipolar lobes is derived to be $\tau_\mathrm{k,bip} = 18,200 \pm 2,200$~yr.

As for the bipolar knots, the image and PV diagrams confirm the presence of at least two pairs labeled NE1, NE2, SW1, and SW2 in Fig.~\ref{fig:slits}. 
Since the inclination and expansion velocity of fast knots or jets are degenerate, the bipolar knots in the Necklace have been assumed to be tilted along a line that keeps an angle with the line of sight as the orthogonal axis of the ring (30$^\circ$) and the symmetry axis of the bipolar structure (46$^\circ$). 
These different assumptions result in the expansion velocities, distance to the CSPN, and kinematical ages listed in Table 1. 
According to the values listed there, the outer knots along an axis orthogonal to the ring would be younger than the bipolar structure, i.e., they should have gone across the bipolar lobes. 
This interaction should have resulted in a disrupted morphology and shocked excitation with bright [N~{\sc ii}] emission of the bipolar lobes, as in the case of NGC\,6778, where outflows younger than the bipolar nebula disrupted it \citep{Guerrero+2012}. 
This is not observed in the Necklace Nebulae, thus we
favor an inclination angle similar to that of the bipolar lobes. 
This implies that the outer knots and the ring are misaligned, in agreement with the different alignment seen in direct images. 
The angular true distances for the NE1 and NE2 knots were determined to be $67\pm2$ and $76\pm2$ arcsec, respectively, corresponding to physical distances from the CSPN of $1.50\pm0.05$~pc and $1.70\pm0.05$~pc. Their expansion velocities resulted in $58\pm5$ and $60\pm5$ km~s$^{-1}$. 
Meanwhile the angular true distances from the CSPN of the SW1 and SW2 knots are $67\pm2$ and $78\pm2$ arcsec, respectively, implying physical distances of $1.50\pm0.05$~pc and $1.74\pm0.05$~pc. 
Their expansion velocities resulted in $68\pm5$~km~s$^{-1}$ and $70\pm5$ km~s$^{-1}$ for the SW1 and SW2 knots, respectively.
The distances of the NE1-SW1 and NE2-SW2 bipolar knots are within their uncertainties and can thus considered to be similar. 
This is also the case for their expansion velocities, with average systemic velocities of $+5\pm5$ km~s$^{-1}$ for both the NE1-SW1 and NE2-SW2 pairs of knots.  
Given the uncertainty of $\pm3$ km~s$^{-1}$ of the systemic velocity of the Necklace Nebula based on the kinematics of the knots in the ring, the bipolar knots share the same systemic velocity. 

The physical properties of the low-ionization clumps, including their kinematical ages, are summarized in Table~\ref{tab:kin}. We find that the outermost clumps (NE2 and SW2) possess larger kinematical ages than the more internal structures (NE1 and SW1). Assuming that NE1--SW1 and NE2--SW2 represent symmetric pairs resulting from distinct ejection events, we can estimate their mean kinematical ages. 
For the NE1--SW1 pair, we derive a weighted mean kinematical age of $\tau_{\mathrm{k1}} = 22,900 \pm 1,900$~yr, which increases to $\tau_{\mathrm{k2}} = 25,500 \pm 2,000$~yr the NE2--SW2 pair of knots.

Fig.~\ref{fig:MO} presents a global view of the model, were we highlight the main structures: the disrupted ring, the bipolar structure, and the two pairs of jet-like features. 
The synthetic PVs obtained from this model are compared with the observations in the bottom panels of Fig.~\ref{fig:N2} and \ref{fig:O}. 

To provide a more rigorous assessment of the morpho-kinematic fit, the observed PV diagrams are overplotted as contours directly onto the synthetic \textsc{shape} models in Fig.~\ref{fig:A2} of Appendix~\ref{app:PV}. 
These figures show that the synthetic emission produced by the model of bipolar knots misaligned with the axis orthogonal to the equatorial ring successfully reproduces the overall spatial and kinematic structure of the observed PV diagrams. 
Some minor differences of the velocity of the northern knot in Slit A and spatial extent of the southern knot in Slit B remains, but these are not unexpected, as \textsc{shape} provides a macroscopic and idealized representation of the nebular geometry and does not account for local turbulent motions, density inhomogeneities, or departures from the adopted symmetric expansion law. 
Nevertheless, the overall agreement between the model and the observations across multiple independent slit positions provides support for a genuine misalignment between the equatorial ring and the bipolar outflows.

\section{Discussion and concluding remarks}
\label{sec:summary}

We have presented a detailed morpho-kinematic study of the Necklace Nebula based on spatially resolved spectroscopy and 3D modeling with the {\sc shape} software. Our results confirm previous findings by \citet{Corradi11} that the nebula is dominated by a dense equatorial ring, accompanied by bipolar outflows whose geometry and kinematics can be consistently reproduced with relatively simple, axi-symmetric structures. The synthetic images and PV diagrams derived from our model provide a good match to the observations, lending confidence to the inferred 3D configuration.

The kinematical ages derived in this study (see Table \ref{tab:kin}) reveal a complex, episodic mass-loss history. The outermost clumps (NE2 and SW2) possess the largest kinematical ages ($\sim$25,500 yr), followed by the internal structures (NE1 and SW1, $\sim$22,900 yr) and the bipolar barrel ($\sim$18,200 yr). The fragmented equatorial ring is the youngest major structure, with an age of approximately 5,090 yr. This chronology indicates that the jets were launched significantly before the common envelope phase, likely during the late AGB phase of the primary. 

One of the key outcomes of this work is the determination of the inclination angle of the bipolar structures, which we find to be $46\pm2^{\circ}$. This value is significantly different from configurations in which the bipolar axis lies close to either the line of sight or the plane of the sky, and it is crucial for correctly interpreting the observed velocity field. The sensitivity of the synthetic PV diagrams to relatively small changes in inclination highlights the importance of 3D modeling when disentangling projection effects in complex PNe such as the Necklace Nebula. Comparisons between the observations and a model in which the bipolar outflows are perfectly aligned with the symmetry axis of the disrupted ring suggest that the bipolar structures are not simply extensions of the equatorial ring, but instead trace a distinct mass-ejection event.

The slight misalignment between the barrel-like bipolar structure and the equatorial ring may indicate either precession of the ejection axis or a temporal change in the dominant mass-loss geometry \citep[see][and references therein]{Ondratschek2022,Vetter2025}. Such behavior is naturally expected in post-CE systems, where the interaction between the stellar cores and the envelope can lead to rapidly evolving outflow conditions \citep[see for instance the case of the bending outflows in NGC\,6337, also misaligned with its toroidal main structure,][]{GarciaDiaz+2009}.
However, the bipolar knots have larger kinematic ages than that estimated for the ring-like structure (see Table~\ref{tab:kin}) indicating that they might have formed by a process preceding the CE phase, as suggested by previous works.  

Accretion in a pre-CE phase might take place through a wind-accretion mechanism \citep[see][and references therein]{TejedaToala2025,Toala2026} or most likely close to the Roche Lobe overflow phase \citep[e.g.,][]{JuarezGarcia2025}. Nevertheless, the formation of an accretion disc around the secondary and the production of bipolar structures previous to the CE phase is likely. 

The misalignment between the bipolar jets direction and the disrupted ring-like structure might lie in the geometry and evolution of the magnetic field responsible for launching the jets. As shown by \citet{Tocknell2014}, the formation of jets requires accretion discs threaded by magnetic fields, whose strength and configuration regulate both angular momentum transport and the collimation of the outflows. If the large-scale poloidal magnetic field is not perfectly aligned with the binary angular momentum axis, the resulting jets will be launched along a tilted direction. Whilst the toroidal structure formed due to the common envelope phase will be coplanar to the binary's orbital plane, which seems to be the case of the disrupted ring-like structure in the Necklace Nebula.
Furthermore, any reorientation or instability in the accretion flow may lead to a misalignment between the disc and the orbital plane. Since the magnetic field is anchored to the disc, such variations would be directly imprinted on the jet axis. In this context, the observed inclination of the polar structures can be interpreted as a natural consequence of a time-dependent configuration of the magnetic field.

These results suggest that magnetic fields not only drive and collimate the outflows, but also play a key role in determining their orientation, providing a physically consistent explanation for the observed deviation of the polar axis from perpendicularity with respect to the equatorial plane. 
The interaction between the nebula and the interstellar medium (ISM) can be envisaged as an alternative explanation for the outflow-ring misalignment. 
Indeed the Gaia EDR3 proper motion of the central star ($\mu_\alpha = -1.321~{\rm mas~yr^{-1}}$ and $\mu_\delta = -3.942~{\rm mas~yr^{-1}}$) implies a considerable tangential velocity ($\approx93$ km~s$^{-1}$ at the distance of 4.6 kpc).  
As the direction of the motion is mostly towards the South (PA $\approx$198$^\circ$), the resulting ram pressure, in conjunction with the relatively low mass of the jets and knots, would make the leading Southern knots closer to the central star than the trailing Northern knots, which is not observed. 
The reason for the negligible interaction with the ISM could be its low density at the height of $\approx250$ pc over the Galactic Plane of the Necklace Nebula. 

In a broader context, this study adds to the growing body of evidence that complex morphologies in PNe are often the result of multiple, distinct mass-loss episodes rather than a single, homologous ejection. The Necklace Nebula exemplifies how high-quality kinematic data combined with 3D modeling can be used to reconstruct this evolutionary history and to place meaningful constraints on the role of binarity in late stellar evolution.

In conclusion, the Necklace Nebula exemplifies how complex morphologies in PNe result from multiple distinct mass-loss episodes rather than a single ejection. Our morpho-kinematic model provides a coherent 3D picture that successfully reproduces both morphology and kinematics. Future high-resolution observations in other wavelengths and tailored hydrodynamical simulations will be essential to further explore the magnetic and binary interactions operating in this remarkable post-CE system.

\section*{ACKNOWLEDGEMENTS}

The authors thank an anonymous referee for comments and suggestions that improve the presentation of the results. E.S. thanks UNAM DGAPA for funding as a postdoctoral researcher. E.S. and J.A.T. thank the support from the UNAM PAPIIT project IN102324. M.A.G. acknowledges financial support from grants CEX2021-001131-S funded by MCIN/AEI/10.13039/501100011033 and PID2022-142925NB-I00 from the Spanish Ministerio de Ciencia, Innovaci\'on y Universidades (MCIU) cofunded with FEDER funds. L.S. acknowledges support from the UNAM PAPIIT project IN107625. E.S. and G.R-L acknowledges support from SECIHTI, Mexico (grant CBF-2026-843). 
This work has made extensive use of NASA's Astrophysics Data System (ADS). 

The completion of this project was achieved thanks to the computing resources of IRyA-UNAM, which are developed and maintained by the personnel of the institute's Computing Department, Daniel Díaz-González, Miguel Espejel, Alfonso Ginori González, and Gilberto Zavala.

We thank the daytime and night support staff at the OAN-SPM for facilitating and helping to obtain our observations, especially Francisco Guillen and Hortensia Riesgo, our telescope operators, for their assistance during observations, as well as the CATT for time allocation. This article is based on observations carried out at the Observatorio Astronómico Nacional on the Sierra San Pedro Mártir (OAN-SPM), Baja California, México. 

\section*{Data Availability}

The data underlying this work are available in the article. The data files will be shared on reasonable request to the corresponding author.



\bibliographystyle{mnras}
\bibliography{rmaa} 




\appendix

\section{Other PV diagrams}
\label{app:PV}

\begin{figure*}
    \centering
    \includegraphics[width=1.0\linewidth]{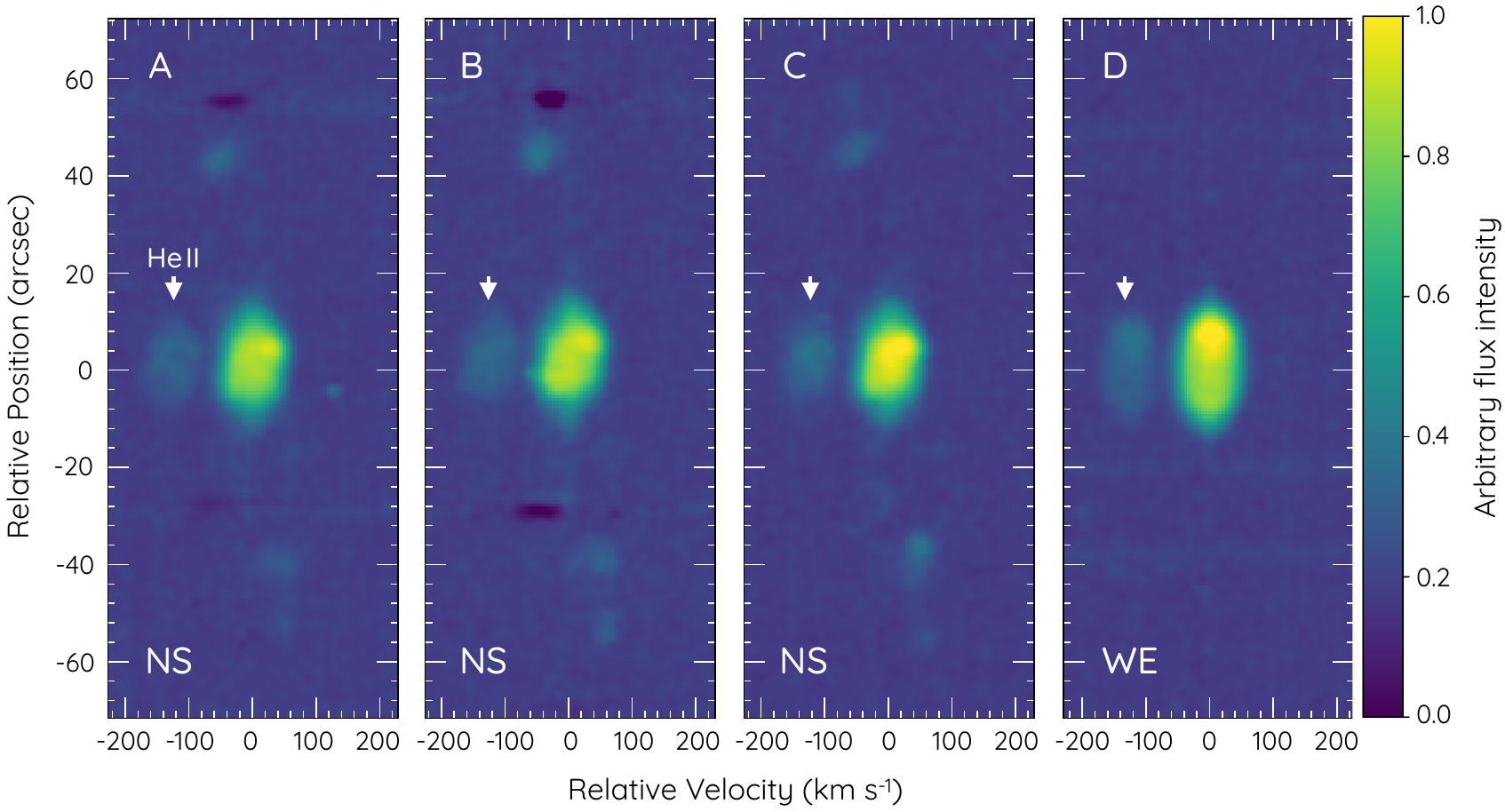}\\[-15pt]
	\includegraphics[width=1.0\linewidth]{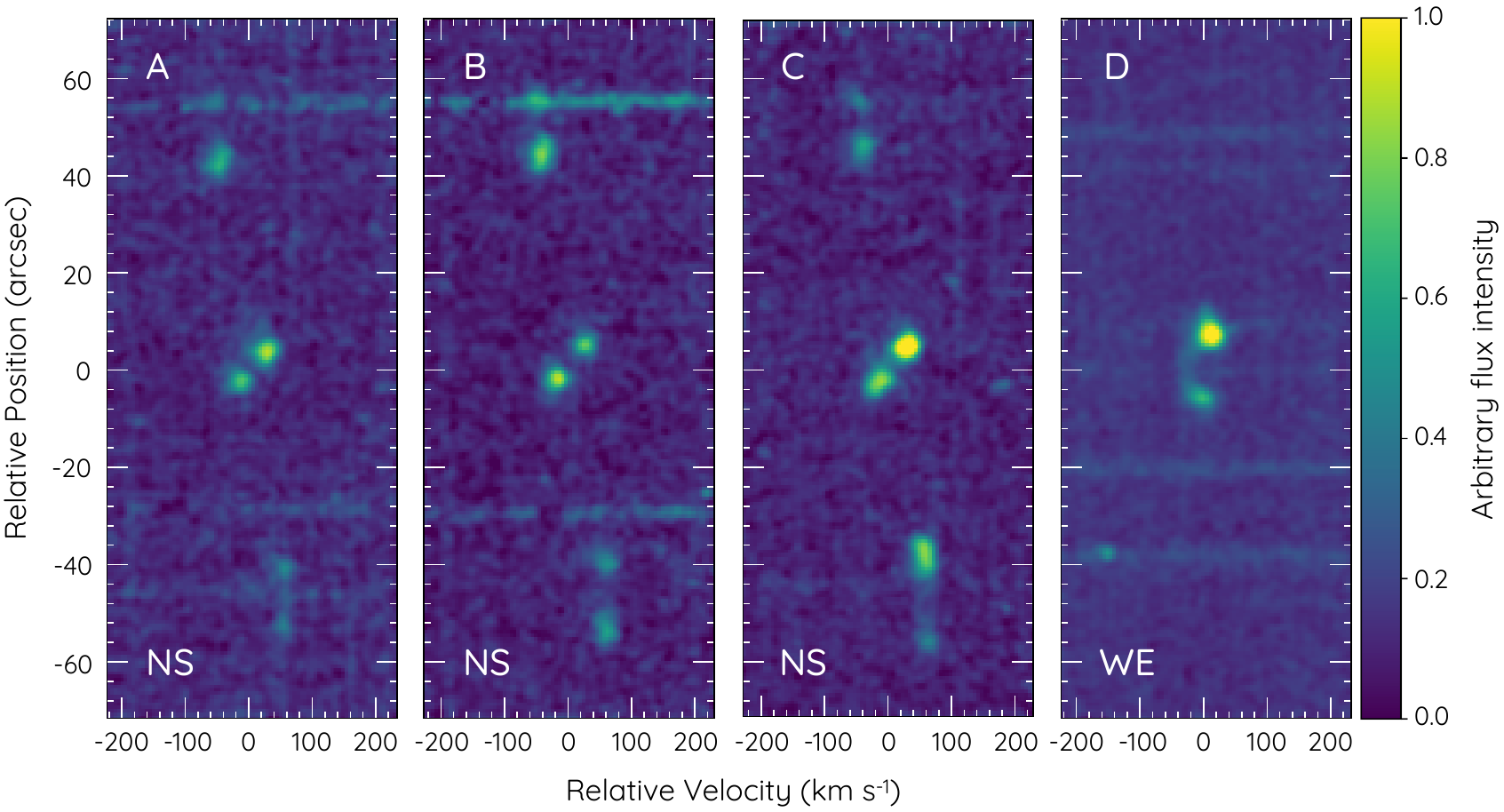}
    \caption{PV diagrams of slits A, B, C, and D extracted from H$\alpha$ (top) and [N~{\sc ii}] $\lambda6548$ \AA, (bottom) emission lines. The presence of the He~{\sc ii} emission in the top panel is shown with a white arrow.}
    \label{fig:N1}
\end{figure*}

Fig.~\ref{fig:N1} presents the PV diagrams centered on the H$\alpha$ (top) and [N~{\sc ii}] 6548~\AA~ (bottom) emission lines for comparison with those of Fig.~~\ref{fig:N2} and \ref{fig:O}. Note that the PV diagrams of the H$\alpha$ emission unveils the presence of the He~{\sc ii} at 6560~\AA.

\begin{figure*}
    \centering
    \includegraphics[width=1.0\linewidth]{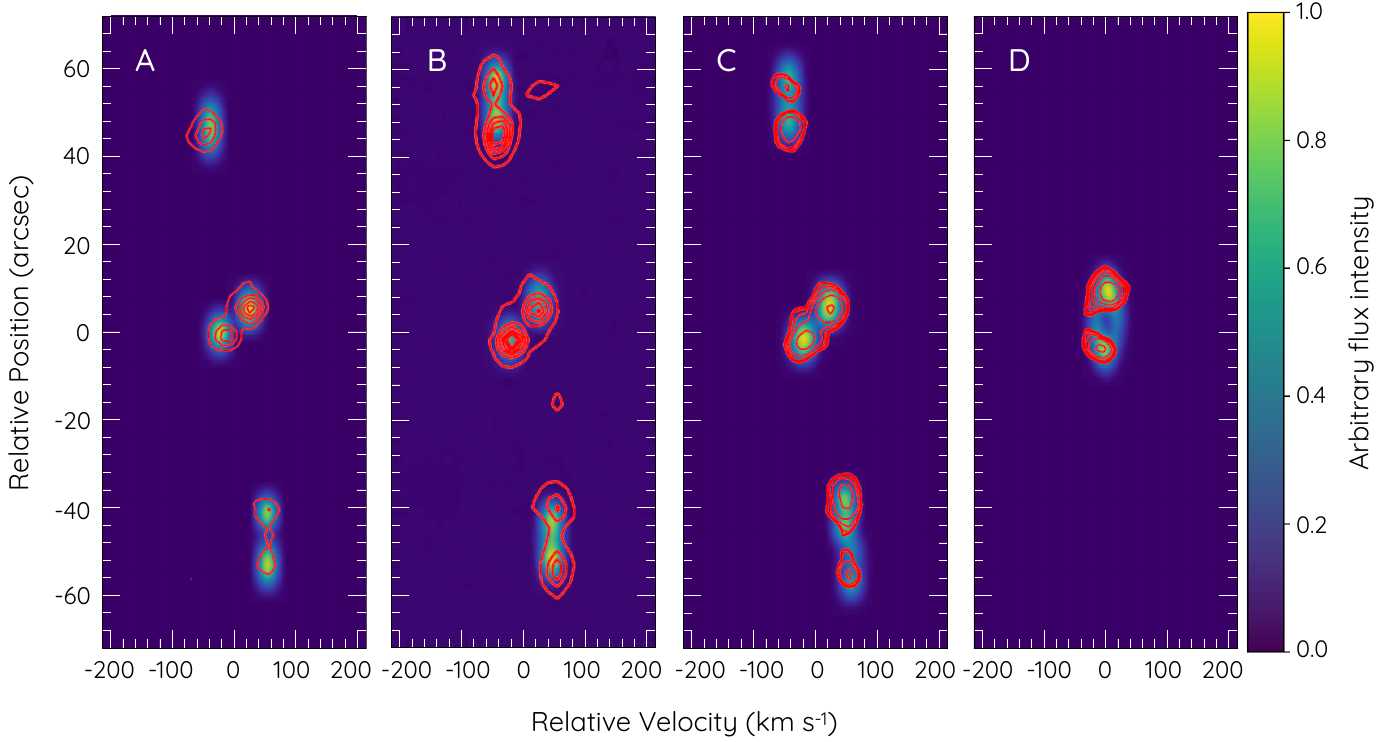}
    \caption{The observational PV [N~{\sc ii}] $\lambda6584$ \AA~ data are overplotted as contours on the synthetic spectra obtained from our best {\sc shape} model to illustrate the precision of the morpho-kinematic fit. This model accounts for a misalignment between the symmetry axis of the disrupted ring and the bipolar ejections.}
    \label{fig:A2}
\end{figure*}

Additionally, to further verify the uniqueness of our best-fitting model, we present in this appendix an extended view of the multi-slit PV arrays with the full observational contour overlays for Slits A, B, C, and D in the [N~{\sc ii}] $\lambda6584$~\AA\ emission line (Fig.~\ref{fig:A2}). These contours represent the relative intensity levels of the observed emission, tracing the exact kinematic paths of the knots and the equatorial ring. The close match between the localized synthetic emission peaks and the cores of the observational contours highlights the high reliability of the derived inclination angles and expansion velocities, confirming that alternative aligned configurations fail to reproduce the multi-slit spatial-velocity distribution.

\bsp	
\label{lastpage}
\end{document}